# Ballistic Thermal Transport at Sub-10 nm Laser-Induced Hot Spots in GaN Crystal

Dezhao Huang[a, #], Qiangsheng Sun[a, #], Zeyu Liu[b], Shen Xu[c], Ronggui Yang[d], Yanan Yue[a, *]

a. School of Power and Mechanical Engineering, Wuhan University, Wuhan, Hubei, 430072, China
b. Department of Applied Physics, School of Physics and Electronics, Hunan University, Changsha, Hunan, 410082, China
c. School of Mechanical and Automotive Engineering, Shanghai University of Engineering Science, Shanghai, 201620, China
d. School of Energy and Power Engineering, Huazhong University of Science and Technology, Wuhan 430074, China

[#] D.H. and Q.S. contributed equally to this work.

*Corresponding authors: Yanan Yue, Email: yyue@whu.edu.cn

# ABSTRACT

Gallium nitride (GaN) is a typical wide-bandgap semiconductor with a critical role in a wide range of electronic applications. Ballistic thermal transport at nanoscale hotspots will greatly reduce the performance of a device when its characteristic length reaches the nanometer scale, due to heat dissipation. In this work, we developed a tip-enhanced Raman thermometry approach to study ballistic thermal transport within the range of 10 nm in GaN, simultaneously achieving laser heating and measuring the local temperature. The Raman results showed that the temperature increase from an Au-coated tip-focused hotspot was up to two times higher (40 K) than that in a bare tip-focused region (20 K). To further investigate the possible mechanisms behind this temperature difference, we performed electromagnetic simulations to generate a highly focused heating field, and observed a highly localized optical penetration, within a range of 10 nm. The phonon mean free path (MFP) of the GaN substrate could thus be determined by comparing the numerical simulation results with the experimentally measured temperature increase which was in good agreement with the average MFP weighted by the mode-specific thermal conductivity, as calculated from first-principles simulations. Our results demonstrate that the phonon MFP of a material can be rapidly predicted through a combination of experiments and simulations, which can find wide application in the thermal management of GaN-based electronics.

# 1. Introduction

Gallium nitride (GaN) is a third-generation wide-bandgap semiconductor, with a promising potential in optoelectronics[1,2] and next-generation power electronics[3–6], owing to its intrinsic ability[7] to withstand ultrahigh current and power densities. However, the design of high-power GaN devices is strongly affected by the self-heating[8,9], and accurate thermal management is needed to control the temperature of such devices. Fourier's law describes heat diffusion when the characteristic length scale of thermal transport is much longer than the phonon mean free path (MFP). However, if the size of a hotspot is small enough, it will induce nonlocal (also called quasi-ballistic) transport phenomena[10,11]. Then, non-diffusive thermal energy phonon carriers will travel from the source before experiencing collisions, and Fourier's law is no longer adequate to describe thermal transport at the nanoscale[12].

Various experiments have been carried out to study quasi-ballistic transport phenomena. Goodson et al.[13] used heating and electrical resistance thermometry to provide experimental evidence for the unexpected temperature increase resulting from ballistic phonon transport. In another experiment, Cahill et al.[14] linked the quasi-ballistic thermal transport to variations in thermal conductivity with the modulation frequency in time-domain thermoreflectance (TDTR) experiments. Tests on $MoS_2$ thin films[15] showed that more than half of the heat was carried by phonons with MFP longer than 200 nm, which exceeded the kinetic theory estimation by nearly two orders of magnitude. The thermal conductivity dependence on the length of silicon nanowires[16] exhibits a transition from semi-ballistic thermal phonon transport at 4 K to fully diffusive transport at room temperature. The energy distribution of heat carriers[17] has also been probed using ultrafast optical spectroscopy. Balandin et al.[18] showed that the experimentally observed linear decrease of the thermal conductivity can be explained by the enhanced phonon relaxation on Si dopants. Minnich et al.[19] used transient thermoreflectance experiments to measure the phonon MFP and observed a dependence of the thermal conductivity on the laser diameter. A previous study found an almost threefold decrease in energy transport from a nanoscale heat source compared with the predictions based on Fourier's law[20]; the authors demonstrated that Fourier's law can still be used if corrected with an additional size-dependent resistance accounting for ballistic effects.

In recent years, theoretical studies[21] have also predicted that the large reduction in thermal conductivity of graphene oxide is due to the significantly enhanced phonon scattering induced by oxygen defects. Chen et al.[22] used the phonon Boltzmann transport equation (BTE) to show that the effective thermal conductivity of superlattices in the perpendicular direction is generally controlled by phonon transport within each layer; moreover, measurement of the thermal boundary resistance between different layers showed[23] that the ballistic-diffusive equations are a better approximation to heat conduction at the nanoscale than Fourier's law. Because solving the BTE and performing brute-force density functional theory (DFT) simulations can be computationally expensive and time-consuming, approximate treatments are necessary to deal with ballistic thermal transport phenomena. Huang et al.[24] found that no clear experimental data strongly supported ballistic thermal conduction of Si or Ge at room temperature. However, experiments on homogeneously alloyed nanowires provided clear evidence for ballistic thermal conduction over several micrometers at room temperature. An expression describing energy transport in both ballistic and diffusive regimes has also been proposed; for single-walled carbon nanotubes in the ballistic-diffusive regime, the thermal conductivity shows a $L^{\alpha}$ dependence.

Raman thermometry is a technique[25–28] used to measure local temperatures with high spatial and temporal resolution and fast acquisition time. Near-field thermal measurements based on tip-enhanced Raman spectroscopy represent a non-contact and reliable optical method to investigate the thermal response of a material. Balandin[29,30] introduced Raman thermometry to measure the thermal conductivity of single-layer graphene; this technique was further applied to a wide range of 2D materials[31–33]. Deng et al.[26] used an ultrafast thermal probing method to measure the reduction in thermal conductivity due to size effects, and successfully measured the temperature and thermal stress in the confined apex region of a nanotip. Despite these prior studies, a comprehensive understanding of ballistic thermal transport in GaN is still lacking, and further studies are needed to probe the characteristic length dependence of the thermal conductivity. Hence, in this study we use Raman thermometry for the direct and accurate measurement of tip heating-induced ballistic thermal transport at sub-10 nm resolution in GaN. We investigate ballistic effects in GaN hotspots through a combination of optical experiments, electromagnetic simulations, and DFT calculations. Our results can have important implications for overcoming issues associated with local hotspots, in order to improve the thermal performance of devices based on GaN.

## 2. Experimental Methods

In the present tip-enhanced Raman thermometry experiments, a Raman laser was employed as a heating source and focused on the gold-coated nanotip of an atomic force microscopy (AFM) instrument. The focused laser beam will generate a highly localized heated region under the gold-coated nanotip. A silicon AFM tip (ScanSens, CSG01 series model) was coated with a 20 nm-thick gold layer with a ~ 35 nm curvature radius. A bulk α-GaN sample with 4.23 × 2.19 × 0.64 cm$^3$ size was grown by the ammonothermal method based on simple chemical reactions[34]. The fabrication process of the GaN substrate was monitored *in situ* through a Staib electron gun to ensure its high quality. The GaN substrate was first examined by Raman spectroscopy, as shown in the bottom-right panel of **Figure 1**. The incident laser beam was focused along the normal direction of the grown GaN film; $E_1$(TO) and $A_1$(LO) were the apparent active phonon modes of GaN in this configuration. The shift in the $E_1$(TO) phonon mode was used to quantify temperature increases, due to its high peak intensity. A typical GaN Raman spectrum displayed a characteristic peak centered around 569.8 cm$^{-1}$ at room temperature; this peak showed a gradual red shift with increasing temperature, as illustrated in the bottom-right inset of **Figure 1**.

A micro-Raman spectrometer was used to determine local temperature increases,

based on the linear relationship between the temperature and the Raman peak frequency of the material. The detailed setup of the optical alignment for thermal sensing using tip-enhanced Raman thermometry is presented in **Figure 1**. The sample was fixed on a substrate holder attached to a three-axis translation stage, making it possible to precisely move the nanotip in the narrow target area. The laser spot was ~ 60 μm in diameter at the focal level in the $z$ direction. The location of the nanotip in the $x$−$y$ plane was adjusted to focus the laser on the apex of the tip and heat the GaN substrate.

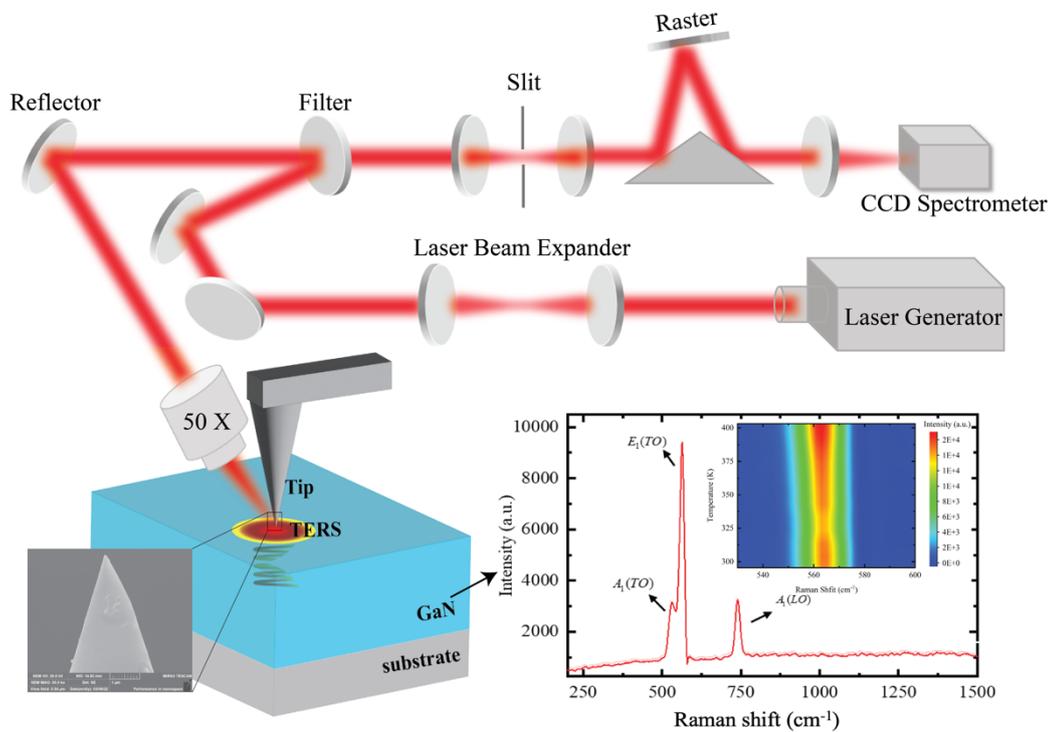

Figure 1. Schematic diagram of optical alignment for thermal sensing. The excited Raman and Rayleigh scattering signals are received by the spectrometer through the same optical path. The incident laser beam at 532 nm crosses the tip and heats the GaN substrate through a 50× convex lens. The red region represents the heating effect induced by the tip. The spiral shows the optical and heat penetration inside the GaN substrate. The bottom-left panel shows a SEM image of the AFM probe tip with a radius of 10 nm and a half-cone angle of 20°. The bottom-right panel shows the Raman spectrum of the GaN substrate, with the inset displaying the red shifts of the GaN Raman peaks with increasing temperature.

Additional irradiations on the cantilever or the tip base should be avoided. The exact location of the laser-irradiated tip was determined by carefully moving the tip; the intensity of the Rayleigh scattering signal was used to establish whether the sample was within the irradiated region. For simplicity, all adjustments were made by referring to a coordinate system established with a resolution of 5 μm. The inset in the bottom-left panel of **Figure 1** displays a scanning electron microscopy (SEM) image of the tip, showing the probe structure used in the subsequent tip-enhanced experiments. The tip was first moved along the direction of the cantilever (*x*-axis) while keeping its position unchanged in the other two directions. When the strongest Rayleigh scattering intensity began to appear, the laser was focused on the tip center. Then, we adjusted the tip along the *y* axis and placed it at the spot where the Raman signal was just detected. The spectrometer and laser generator were connected to a computer and controlled *via* a pre-installed software. The feedback signals were received and processed using a program implemented in the support software.

## 3. Results and Discussion
### 3.1 Raman Calibration Results

Detailed calibration analysis of the GaN substrate was carried out to analyze the relationship between temperature and Raman shifts. The sample was placed on a heating stage to heat it at different temperatures and measure the corresponding Raman spectra. **Figure 2a** shows the linear relationship between the Raman shift corresponding to $E_1$(TO) modes and the temperature. The observed linear trend of the Raman shifts of GaN can be attributed to the lattice thermal expansion and contraction due to temperature changes; in particular, the phonon energies shift in response to changes in the equilibrium positions of atoms. Balkanski et al.[35] presented a theoretical model of the Raman frequency dependence on the temperature and pointed out that the frequency and temperature exhibit a linear relationship if the temperature is under 600 K, due to phonon anharmonic effects. This relation can be expressed as $\omega = C(T_\omega - T_0) + \omega_0$, where $\omega$ is the measured Raman shift of a particular mode, $T_\omega$ is the "Raman peak-independent" temperature, $C$ is a calibration constant to be determined, and $\omega_0$ is the Raman shift at room temperature. Its temperature coefficient of the GaN during its calibration was calculated as −0.014 cm$^{-1}$ K$^{-1}$ in the calibration experiment, which was in good agreement with previously reported values[1–3]. The obtained temperature coefficient can be further used to measure local temperature increases in the GaN substrate under a highly focused hotspot introduced by the gold-coated tip upon laser irradiation.

The temperature increase in the GaN heating spot is directly related to the thermal conductivity of the material; as the size of the tip is only 10 μm in diameter in the apex region, the localized electromagnetic field increase induced by surface plasmon resonance will create a hotspot enhancing the Raman signals, as shown in **Figure 2b**. The results reported in the following can be explained in terms of ballistic thermal transport phenomena; this is because the size of the heated region induced in the GaN substrate is close to the size of the tip, and its characteristic length is much shorter than the typical phonon MFP of GaN. In the bulk GaN system, most heat is transported *via* phonons with MFPs of 100–1000 nm at room temperature; thus, a ballistic behavior obeying Landauer's formulation for quantum thermal conduction is expected. After the laser beam is focused on the GaN substrate, heat will be generated on both the uncoated and gold-coated silicon tip systems. However, the gold film covering the tip can enhance light absorption and thus promote ballistic thermal transport.

### 3.2 Raman Probing of GaN Thermal Response

**Figure 2c** displays seven representative Raman spectra corresponding to the bare silicon tip, collected at laser powers ranging from 4.5 to 13.8 mW. As the laser power increases, the Raman-active modes are softened, due to the increased local heating of the GaN substrate originating from the laser irradiation, resulting in a stronger phonon–phonon scattering. It should be noted that a laser excitation power up to 13.8 mW will cause no damage to the sample, while the linear dependence of the GaN Raman shift on the temperature is maintained. If the laser power intensity is decreased to limit the sample heating, this will reduce the signal-to-noise ratio of the Raman peak and affect the comparison. Different laser powers also result in a red shift and higher intensity of the Raman peaks, as shown in **Figure 2d** for the gold-coated silicon tip case, with the plasmon resonance enhancement of the Raman signals providing a better spatial thermal resolution. In detail, the increased Raman peak intensity of the GaN substrate under the gold-coated silicon tip is almost 1.5 times higher than that measured in the bare tip case. The above observation further clarifies the decreasing trend of the GaN

Raman shift with the laser power, as shown in **Figure 2e**.

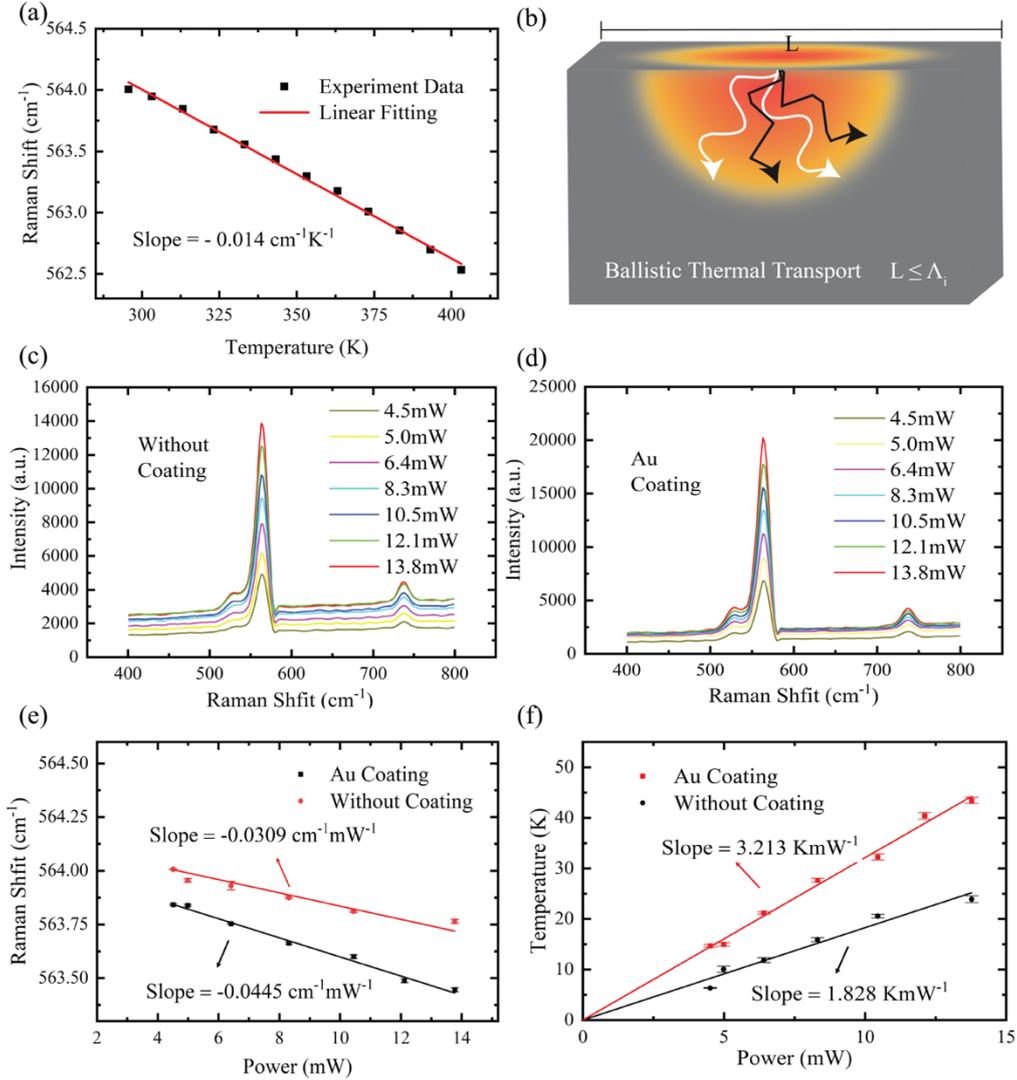

Figure 2. (a) Linear relationship between Raman peak position and temperature; the slope of the linear fit was used to determine the GaN substrate temperature. (b) Ballistic thermal transport regimes. Ballistic phonon thermal transport will occur when the length of the heat source (L) is shorter than the phonon MFP. (c) Selected spectra of GaN under bare silicon tip irradiated with different laser powers. (d) Selected spectra of GaN under gold-coated silicon tip irradiated with different laser powers. (e) Raman shift vs. laser power plot. The slope of the fitting line corresponding to the uncoated tip is almost two times larger than that of the gold-coated tip. (f) Comparison of tip-enhanced temperature increases as a function of laser power with and without Au coating. The gold-coated tip greatly enhances thermal transport under the tip. The error bar is the standard deviation from three different independent experimental measurements.

The dependence of the GaN peak frequency on the laser power in the linear region can be described as $\omega = \chi_p(P_2 - P_1) = \chi_p \Delta P$, where $\chi_p$ is the slope of the Raman shift *vs.* laser power plot and $\Delta P$ is the laser power difference. The fitted coefficients $\chi_p$ for the bare and gold-coated silicon tips are −0.0445 and −0.0309 cm$^{-1}$ mW$^{-1}$, respectively. The decreasing linear power dependence indicates that, for the same laser power, the phonon and thermal energy can be absorbed more efficiently, leading to a higher local temperature. This underlying trend can be extracted using the previous temperature–Raman shift calibration. **Figure 2f** clearly shows that the temperature of the GaN hotspot increases as a function of the laser power, reflecting the enhanced energy absorption by surface plasmon resonance. The temperature increase (~ 40 K) induced by the gold-coated tip is much higher than that (~ 20 K) generated by the bare silicon tip. Moreover, the fitted temperature coefficients for the gold-coated and bare tip cases are 3.213 and 1.828 K·mW$^{-1}$, respectively.

The temperature increase in the uncoated silicon tip case mainly derives from the heating effect of the Raman laser, and can be approximately estimated through the simple heat conduction model[36]. The temperature rise in a small region of a semi-infinite medium under a constant heat flux can be calculated as $\theta_\infty = \frac{2}{\pi} \cdot \frac{q_0 a}{k}$, where $k$ is the thermal conductivity of the material, $q_0$ is the constant heat flux, $a$ is the spot radius, and $\theta_\infty$ is the temperature increase in the small region after thermal equilibrium. The laser spot radius is ~ 0.2 μm and the constant heat flux can be calculated from the heating power (13.8 mW) of the incident laser. The thermal conductivity of single-crystal GaN at room temperature is ~ 253 W/(m·K). The temperature increase predicted from the heat conduction model is 22 K, which is close to that (21 K) estimated from the Raman experiments. Hence, heat transfer in the limited region below the bare tip follows the standard diffusive thermal transport mechanism; however, whether the anomalous temperature increase in the region below the gold-coated tip originates from ballistic thermal transport remains an open question.

### 3.3 Electromagnetic Simulations of Tip–Substrate System

The observed temperature increase in the sub-10 nm region motivated us to simulate the distribution of the electric field and the temperature rise at the tip-enhanced nanoscale heating spot. **Figure 3a** shows the simulated gold-coated tip and GaN substrate system under 532 nm laser irradiation at the incident angle ($\theta$) of 20°. Because the surface plasmon resonance is sensitive to several properties, such as excitation wavelength, material characteristics, polarization of the incident laser, and the laser shape and the careful selection of the input parameters is a crucial task. The model used in the simulation consisted of a silicon tip with a sharp end tangent to a hemisphere and covered by an approximately 20 nm-thick gold coating. A 100 nm-thick silicon substrate was placed normally under the tip. The geometry of the tip was as follows: half taper angle $\theta = 10°$, apex radius $r_1 = 30$ nm, $r_2 = 50$ nm, and length $L = 300$ nm. Maxwell's equations were solved by the finite difference time domain (FDTD) method across a rectangular computational domain with $600 \times 600 \times 800$ nm$^3$ dimensions, containing the tip, substrate, and a vacuum region around the tip/substrate system.

To prevent reflections at outer boundaries, the perfectly matched layer (PML) technique was applied to the domain. A polarized plane wave was irradiated along the *x* direction on the tip at an angle *θ* with respect to the tip axis. The incident electric vector $E_0$ had an amplitude of 1 V/m. At 532 nm (wavelength of the laser used in the experiment), the dielectric functions of gold and silicon are[37] $\varepsilon_{Au}$ = −4.2854 + 2.3292*i* and $\varepsilon_{GaN}$ = 5.8254 −0.4070*i*, respectively. When a polarized laser illuminates the tip, the *z*-component of the polarized electric field drives the free electrons in the near-surface region of gold and confines them to the end of the tip apex. As a result, a strong electric field is generated under the tip apex. Cross-section views of the electric field distributions in the bare tip–substrate and coated tip–substrate systems are shown in **Figures 3b** and **3c**, respectively. Because the tip can be effectively excited using a longitudinal wave, the electric field magnitude around the contact point region of the gold-coated tip is greatly enhanced; in contrast, the other region of the gold-coated tip–substrate system shows a low electric field enhancement. The observed intensification of the electric field may originate from surface plasmon resonance, due to the presence of the thin gold layer on the tip. The electric field enhancement at the junction between tip and substrate can also be seen in **Figure 3c**.

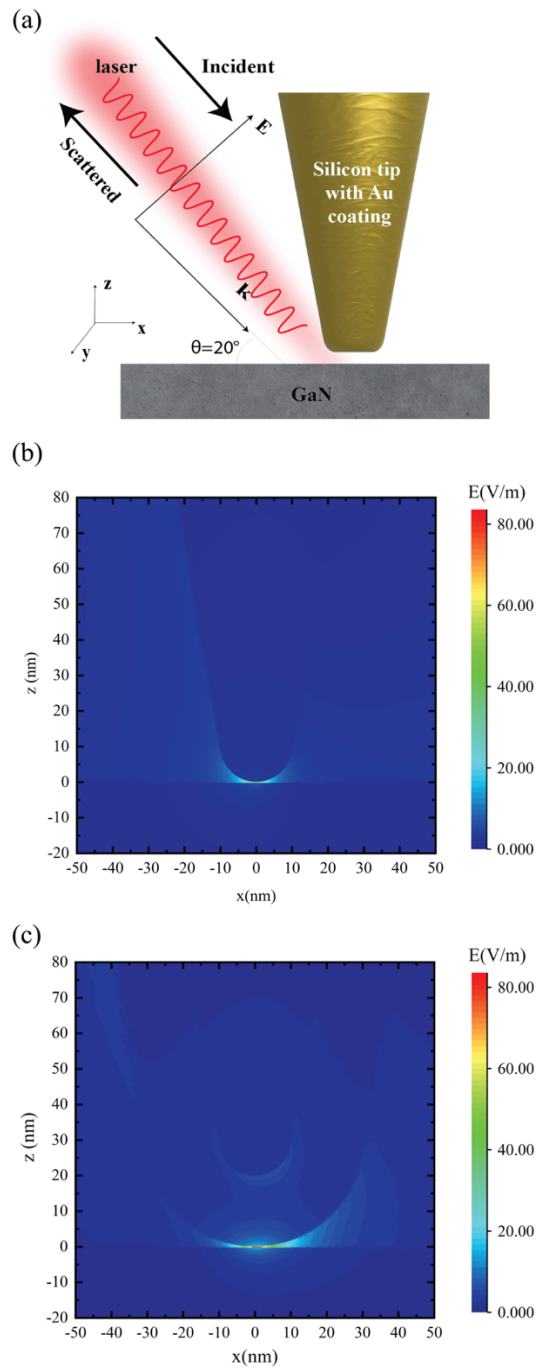

Figure 3. (a) Schematic diagram of tip–substrate simulation model. The tip with a silicon core is coated with a 20 nm-thick gold film. The incident angle of the irradiating laser is 20°. (b,c) Front view of electric field distribution around the bare (b) and gold-coated (c) tip in the $y = 0$ plane. A strongly enhanced electric field is observed around the tip–substrate contact area, with an asymmetric distribution around the tip.

The electric field distributions around the contact point region for the bare tip– and gold-coated tip–substrate systems are shown in **Figures 4a,b**. The enlarged view of the contact point shows that the enhanced electric field only exists in a small region with a radius of less than 10 nm. While most of the electromagnetic wave goes around the tip, a small portion still penetrates into the gold coating, and even into the silicon core. A strong field enhancement is observed near the contact point, both in the tip and the sample; in particular, the electric field increases by 6400 times near the tip region, which is close to the results of previous reports[38]. The field then drops dramatically toward the inside of the tip and the sample. This electrical/optical field in the substrate gives rise to the sub-10 nm heating. The top views of the electric field distributions for the bare and gold-coated tip–substrate systems in **Figures 4c,d** show that the enhanced field in the GaN substrate rapidly declines around 10 nm and gradually decreases above 20 nm. The absorption zone has a nearly hemispherical shape with a diameter of 10 nm, in which the laser radiation is absorbed and converted into thermal energy; the local temperature of the substrate will then show a corresponding increase.

### 3.4 Thermal Analysis in the Sub-10 nm Region

The distributions of the absorbed light and converted thermal energy were calculated with an incident intensity $I_0$ of $4.9 \times 10^6$ W/m$^2$, corresponding to that of the laser available in our lab. In the following, the heat generated in the silicon surface was calculated at this moderate intensity, which is equivalent to $E_0 = 6.1 \times 10^4$ V/m. The relation between intensity and normalized electric field[39] is $I = 0.5n\varepsilon_0 c_0 E^2$. Based on the Poynting's theorem, when light passes through the absorbing medium, the localized heat generation rate $q_{\text{loc}}$ is given by the equation[40,41] $q_{\text{loc}} = 0.5\varepsilon_0\omega\text{Im}[\varepsilon(\omega)]E_{\text{loc}}^2$, where $\varepsilon_0$ is the vacuum permittivity, $\omega$ is the angular frequency of the incident wave, $\text{Im}[\varepsilon(\omega)]$ is the imaginary part of the dielectric function of Si at the incident frequency, and $E_{\text{loc}}$ is the normalized local electric field. **Figures 4c,d** show the distribution of the heat generation rate around the contact point region inside the GaN substrate for the bare and gold-coated tips under irradiation with the 532 nm laser. The hemi-ellipsoidal distribution of the thermal energy density is consistent with the electric field distribution in **Figure 4d**, reaching its maximum of $\sim 6 \times 10^{15}$ W/m$^3$ at the contact point and then dropping dramatically inside the GaN substrate. The asymmetric distribution in the $x$-direction can be attributed to the uneven distribution of the tip-induced optical enhancement, due to the asymmetric illumination relative to the cone axis[42–44]. Thus, a deeper penetration into the GaN substrate results in a more dramatic drop in the heat generation rate. The heat penetration depth along the $z$ axis is around 2 nm, while the lateral penetration depth along the $x$ axis is around 6 nm. The lateral and vertical sizes of the optically induced heated region are as small as $\sim 6$ and $\sim 2$ nm, respectively. The energy absorption concentrated near the tip–sample contact point will be dissipated inside the GaN substrate.

In order to calculate the temperature increase in the heating region of the GaN substrate, the heat transfer process can be described by the traditional heat conduction model for a small particle embedded in a medium. The thermal resistance of the GaN substrate, calculated as $1/(2\pi r k_{\text{eff}})$, where $r$ is the radius of the heated spot (5 nm) and $k_{\text{eff}}$ is the local thermal conductivity, is $1.77 \times 10^5$ K/W. Moreover, the thermal contact resistance between tip and substrate, calculated through molecular dynamics simulations (see supplemental material for more information), is $7.04 \times 10^8$ K/W; although this allows very little heat flow from the substrate to the tip, a large amount of heat flows from the heating region to other regions of the GaN substrate. As shown in **Figures 4e,f**, the length of the heated region is much shorter than the phonon MFP of the GaN substrate; hence, such nanoscale hotspot will not follow the classical Fourier's law of heat conduction in the diffusive regime, due to the failure of the gold coated tip–substrate system to reach thermal equilibrium in such small space. The size of the generated hotspot is around 10 nm, and the induced temperature increase with the 13.8 mW irradiating laser is higher than 330 K. Taken together, the present experimental and simulation results shows that ballistic thermal transport can induce the highly localized heating observed in the GaN substrate.

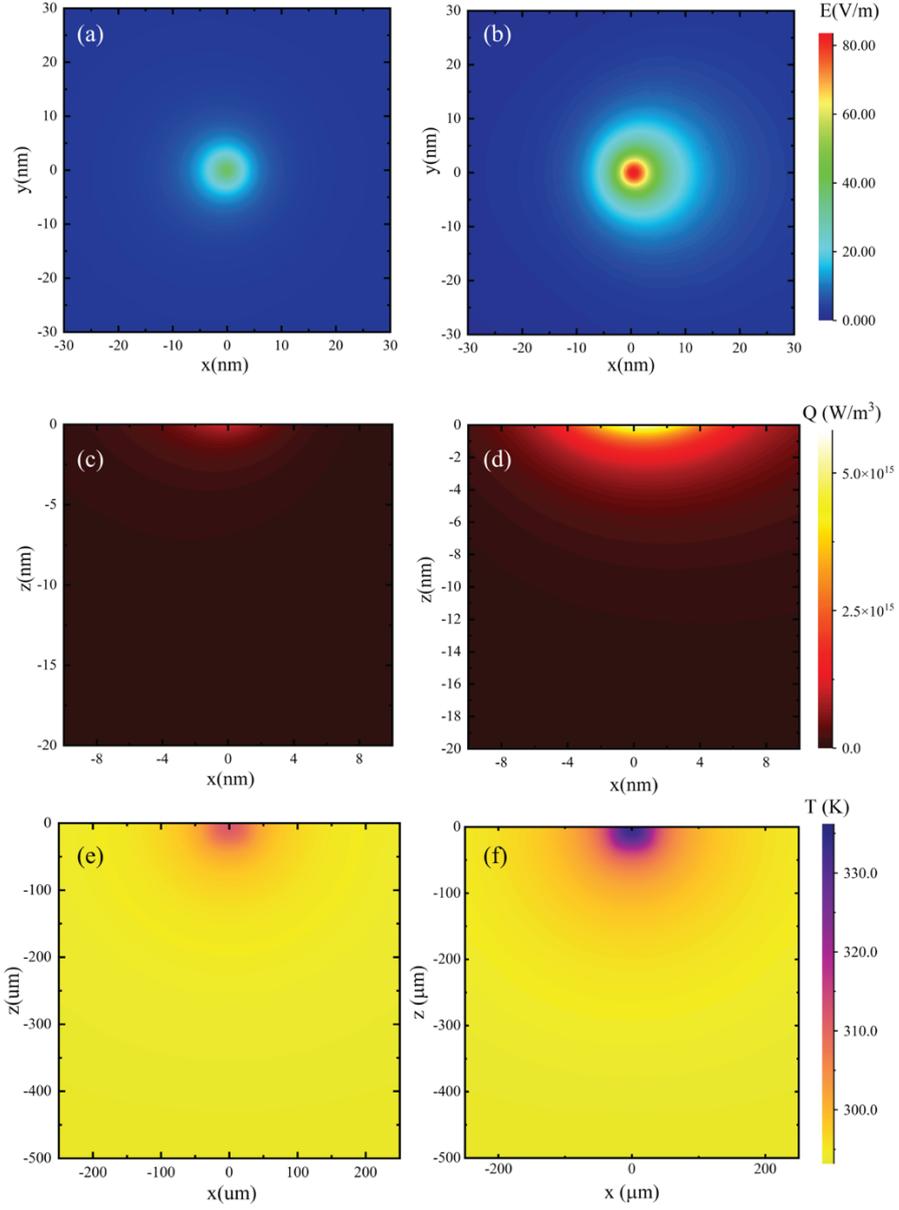

Figure 4. (a,b) Top view of electric field distribution in GaN substrate for bare (a) and gold-coated (b) tip. (c,d) Thermal energy distribution inside GaN sample induced by bare (c) and gold-coated (d) tip. (e,f) Temperature distribution of bare- and gold-coated tip-enhanced heating. The thermal energy leads to a highly localized temperature increase in the GaN sample, and gradually decreases as the penetration depth increases.

## 3.5 MFPs of GaN at Sub-10 nm Hotspot

As shown in **Figure 5a**, a predictive model of phonon MFPs was established by systematically varying the phonon MFP and comparing the corresponding changes in the temperature distribution in the heated region with the experimental values. The phonon MFP of a new material can thus be predicted based on the simulated temperature changes in the heated region, as shown in **Figure 5b**. In general, higher temperature leads to smaller phonon MFP, which is intuitively true considering that higher temperature means stronger phonon-phonon scattering. Note that the temperature measured through Raman experiments includes only the average temperature increase from the heated region, and the temperature estimated from the simulation shows a smooth distribution in this region. The simulated temperature increase was spatially averaged to compare it with the experimentally measured increase. The combination of iterative EM simulation and quasi-ballistic experiments enables the rapid determination of the phonon MFPs *via* a simple search algorithm. According to our simulations, the phonon MFP of GaN is around 797 nm long when the irradiating laser power is 4.5 mW. Various computational approaches have shown that the main phonons contributing to the thermal conductivity of bulk GaN have MFPs of 200–1000 nm[45,46].

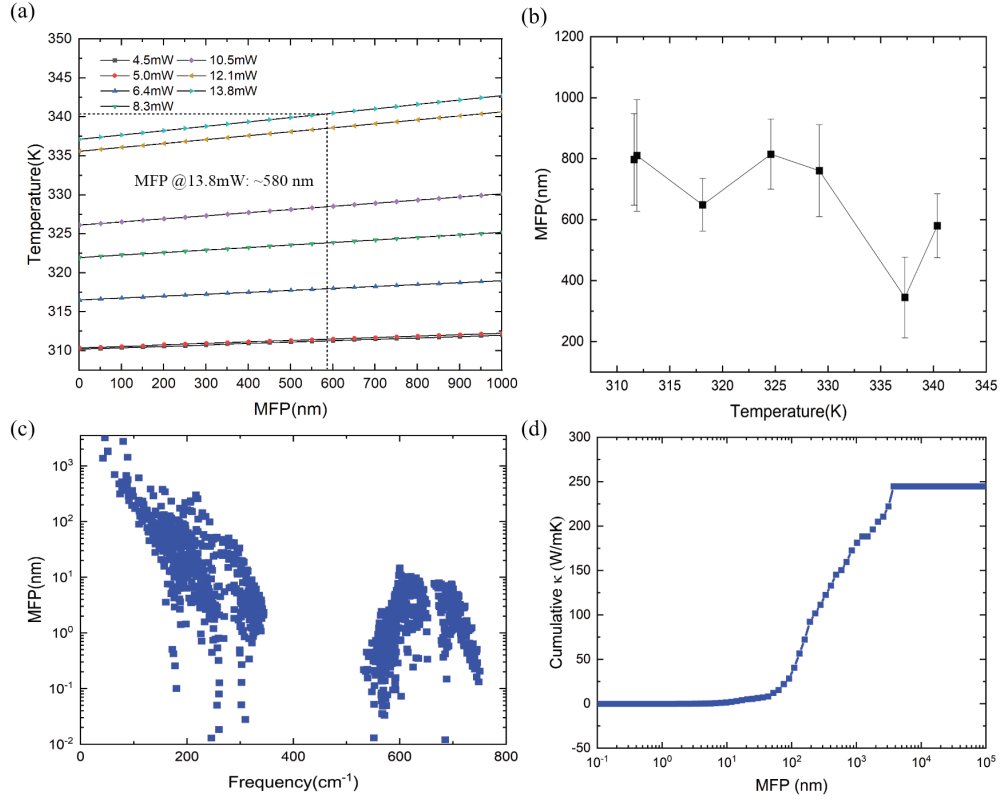

Figure 5. (a) Temperature in the heated region as a function of MFP length, evaluated through electromagnetic field simulations. Unknown MFPs can then be predicted based on the experimentally observed temperature. (b) Predicted MFP length as a function of the heated region temperature in the GaN substrate and the error bar is obtained through three Raman experiments. (c) BTE-calculated MFP length of each phonon mode in the first Brillouin zone of GaN at 300 K. (d) Cumulative thermal conductivity as a function of phonon MFP length at 300 K.

The MFPs obtained from the combination of tip-enhanced Raman thermometry and electromagnetic finite element method (FEM) simulations were then compared with the MFP calculated from first-principles simulations, as shown in **Figure 5c**. To verify the calculated MFPs, the phonon BTE was solved numerically with the help of force constants obtained from first-principles simulations. The present first-principles calculations include two different scattering mechanisms: three-phonon and isotopic scattering. An iterative process was used to obtain the converged phonon distribution and thermal conductivity using the ShengBTE package[47] with an uniform 12 × 12 × 12 grid in the first Brillouin zone. A 4 × 4 × 4 *q*-grid was used to calculate harmonic force constants using the density functional perturbation theory scheme implemented in the Quantum Espresso software; third-order force constants were calculated using a 4 × 4 × 4 supercell, *via* a finite difference method with the cutoff set at the eighth nearest neighbor. The isotopic scattering was then calculated based on the natural isotopic distribution of Ga and N.

A mode specific MFP spectrum is calculated, and the MFP weighted by mode specific thermal conductivity is then averaged as 768.6 nm at 300K for the bulk GaN. This is in good agreement with our tip-enhanced Raman thermometry and FEM approach, indicating that this novel method can successfully capture the nature of ballistic transport. Finally, **Figure 5d** shows the cumulative thermal conductivity as a function of the phonon MFP, providing an estimate of the range of the phonon MFPs that mostly contribute to the thermal conductivity. The analysis shows that more than half of the thermal conductivity is contributed by phonons with a mean free path longer than 500 nm, even at room temperature. In addition, only a small portion of the thermal conductivity originates from phonons with mean free paths shorter than 50 nm. Therefore, we expect a significant reduction of the thermal conductivity when the size of the heated region is around 20 nm, along with a large temperature increase inside this limited region.

# Conclusion

In summary, we used the novel tip-enhanced Raman thermometry to observe the quasi-ballistic thermal transport phenomena in GaN. The experimental results are consistent with ballistic thermal transport when the characteristic thermal transport length is shorter than the phonon mean free path. To further investigate the thermal transport mechanism at the nanoscale, we conducted optical electromagnetic simulations of AFM tip–substrate systems to model the highly focused optical and heating field induced by a laser. The simulation results also show a highly localized optical penetration depth within a range of 10 nm. Based on the temperature increases measured experimentally, optical electric field simulations can be used to determine the phonon MFPs of GaN. Consistent with the phonon MFPs of GaN reported in the literature and the averaged MFP weighted by the mode-specific thermal conductivity calculated by first-principles simulations, we successfully carried out the first determination of MFPs combining Raman experiments and numerical simulations. Our analyses reveal the fundamental mechanisms of ballistic thermal transport under a heating tip, which may provide new insights useful for understanding nanotip-enhanced heating and designing more efficient nanofabrication systems.

## ACKNOWLEDGEMENTS

This work is supported by National Natural Science Foundation of China (No. 52076156), the National Key R&D Program of China (No. 2019YFE0119900) and China Postdoctoral Science Foundation (No.2022M712447). Z.L. would like to thank the Fundamental Research Funds for the Central Universities (Grant No.

531118010723). The simulations were supported by the Research Computing Center in Wuhan University and the National Supercomputing Center at Changsha (NSCC), Hunan University.

## COMPETING INTERESTS

The authors declare no conflict of interest.